\documentclass{article}
\usepackage{spconf,amsmath,amssymb,graphicx,hyperref,booktabs}

\makeatletter
\long\def\@makecaption#1#2{\vskip 4pt
 \setbox\@tempboxa\hbox{#1. #2}%
 \ifdim \wd\@tempboxa >\hsize #1. #2\par \else \hbox
to\hsize{\hfil\box\@tempboxa\hfil}\fi}
\makeatother

\let\spconfthebibliography\thebibliography
\renewcommand{\thebibliography}[1]{\spconfthebibliography{#1}%
  \setlength{\itemsep}{0pt}\setlength{\parsep}{0pt}}

\title{LiveSynth: A Streaming Neural Synthesizer \\
for Instrument Cloning and Text-to-Instrument}

\name{Kyungsu Kim$^*$, Yejin Kim$^*$, Kyogu Lee$^{* \dag \ddagger}$}
\address{$^*$ Music and Audio Research Group (MARG)\\
  $^\dag$ Department of Intelligence and Information, Seoul National University\\
  $^\ddagger$ Interdisciplinary Program in Artificial Intelligence, Seoul National University}

\begin{document}
\ninept
\maketitle

\begin{abstract}
Recently, generative neural synthesizers have enabled zero-shot instrument cloning from reference recordings and timbre generation from text prompts.
However, existing systems generate complete audio segments offline, precluding their use as playable instruments. 
We present LiveSynth, a streaming neural synthesizer with flexible timbre control and interactive generation capabilities, including zero-shot instrument cloning and text-to-instrument synthesis.
A feedback-free causal Transformer conditioned on MIDI events and a CLAP timbre embedding outputs continuous audio latents at 100 Hz, which a causal variational autoencoder decodes into audio. 
Frame-wise timbre conditioning enables continuous timbre morphing, while MIDI-conditioning dropout enables autonomous continuation from a performed prefix.
On the NSynth and Slakh datasets, LiveSynth achieved timbre similarity and spectral reconstruction quality comparable to offline baselines.
On an M3 MacBook Air, LiveSynth required 8.8~ms of computation per 10~ms audio frame---an RTF of 0.88---with 10~ms algorithmic latency, enabling real-time interaction with a performer.
\end{abstract}


\section{Introduction}
\label{sec:intro}

\begin{figure*}[t]
  \centering
  \includegraphics[width=0.95\textwidth]{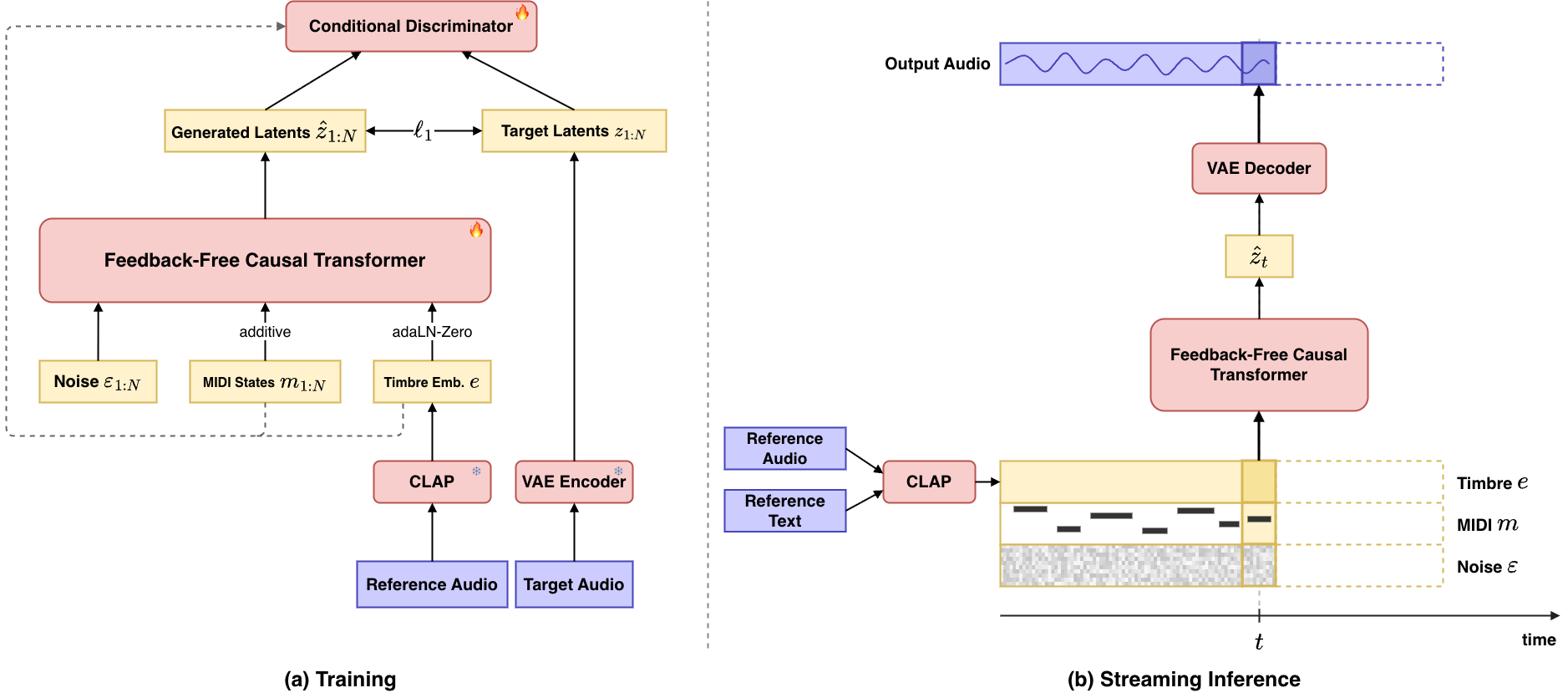}
  \small
\caption{Overview of LiveSynth. 
(a) Conditional adversarial training of the feedback-free continuous-latent generator.
(b) Frame-wise streaming synthesis conditioned on MIDI and audio- or text-derived timbre embeddings, followed by causal VAE decoding.}
  \label{fig:overview}
\end{figure*}

Recent work has introduced a class of models that we refer to as \emph{generative neural synthesizers}: systems that learn a conditional distribution over audio given a symbolic performance.
This class has expanded to include models that accept reference audio or text prompts as timbre conditions, enabling zero-shot instrument cloning, text-to-instrument synthesis, and timbre transfer without target-specific retraining~\cite{tokensynth,ctd,nclmctt,anysynth}.
Conditioning on reference audio or text prompts enables intuitive and flexible control over instrument timbre.
However, existing systems generate complete audio segments offline rather than responding to performance controls in real time, making them unsuitable as playable instruments.

On the other hand, differentiable digital signal processing (DDSP) offers a complementary approach to playable neural synthesis by incorporating differentiable signal-processing priors~\cite{ddsp,hayes2024ddsp}.
Its efficiency supports real-time operation and extensions to polyphonic synthesis~\cite{ddsp_piano}. However, adapting to an unseen timbre generally requires instrument-specific training or optimization~\cite{jonason2022cloning}, leaving streamable zero-shot timbre control unresolved.

We present LiveSynth, a generative neural synthesizer for low-latency streaming with flexible timbre control and interactive generation capabilities, including zero-shot instrument cloning and text-to-instrument synthesis.\footnote{\begin{tabular}[t]{@{}l@{}}Audio examples and source code are available at:\\[-1pt]
\url{https://kyungsukim42.github.io/livesynth-demo/}
\end{tabular}}
Conditioned on streaming MIDI events and frame-wise timbre conditions derived from CLAP~\cite{clap} embeddings, LiveSynth produces continuous audio latents, which a causal audio decoder converts into waveform samples.
LiveSynth builds on the feedback-free continuous sequence modeling framework of LiveBand~\cite{liveband}, without feeding previously generated latents back into the generator.
This design avoids feedback-induced error accumulation while enabling strictly causal inference.
Frame-wise timbre updates enable continuous timbre morphing, while MIDI-condition dropout allows the model to autonomously continue a preceding performance when provided with a null MIDI condition, supporting call-and-response jamming.

Across the NSynth and Slakh datasets, LiveSynth achieved timbre similarity and spectral reconstruction quality comparable to offline baselines.
On an M3 MacBook Air, LiveSynth achieved an RTF of 0.88 with a 10~ms algorithmic latency, satisfying the computational requirements for real-time inference.
Our contributions are as follows.
\textbf{(1)} We introduce a low-latency streaming generative neural synthesizer that supports zero-shot instrument cloning from reference audio and text-to-instrument synthesis within a single model;
\textbf{(2)} We employ frame-wise timbre conditioning to enable continuous timbre morphing during generation;
\textbf{(3)} We apply MIDI-conditioning dropout during training so that the model can continue a preceding performance from a null MIDI condition, enabling call-and-response interaction with a human performer.

\section{Related Work}
\label{sec:related}

Early generative neural synthesizers such as NSynth and Mel2Mel learned timbre representations that enable controllable synthesis and interpolation~\cite{nsynth,mel2mel}.
Alongside this development, DDSP-based synthesizers incorporated differentiable signal-processing priors for efficient real-time operation, although adapting to unseen timbres generally requires target-specific training or optimization~\cite{hayes2024ddsp,jonason2022cloning}.
More recent generative neural synthesizers, including TokenSynth, CTD, and AnySynth, condition generation on reference audio or text prompts, enabling zero-shot instrument cloning and open-ended timbre control~\cite{tokensynth,ctd,anysynth}.

Meanwhile, real-time and interactive music generation has been explored across multiple forms of musical interaction.
Unified Timbre Transfer enables within-stream morphing through time-varying timbre embeddings~\cite{unified_timbre_transfer}.
Live Music Models and LiveBand generate continuous audio streams in real time, with LiveBand focusing on accompaniment conditioned on live audio~\cite{live_music_models,liveband}.
ReaLJam explores responsive human--AI performance by accompanying a musician's input~\cite{realjam}.

\section{Method}
\label{sec:method}

Figure~\ref{fig:overview} illustrates the overall architecture of LiveSynth during training and streaming inference.
Given streaming MIDI events and a timbre condition derived from reference audio or text, LiveSynth generates continuous audio latents, which a causal VAE decoder converts into audio as they are produced.
The generator architecture and conditional adversarial formulation follow the feedback-free continuous sequence modeling framework of LiveBand~\cite{liveband}, in which generated latents are not fed back into the model.

\subsection{Causal Audio Autoencoder}

One direct approach to streaming generative synthesis would be to apply language modeling to quantized neural audio codec tokens~\cite{audiolm,tokensynth}.
However, achieving low latency requires short codec frames and short or zero lookahead, resulting in a high-rate, strictly causal representation.
Our preliminary audio autoencoder experiments showed that these constraints substantially reduced reconstruction and perceptual quality, with latent quantization causing further degradation. 
We therefore omit the quantization layer and adopt the 100~Hz continuous latent space of a causal audio autoencoder.

We train a Gaussian VAE separately and freeze it before training LiveSynth.
Its encoder and decoder are adapted from DAC and Vocos, respectively~\cite{dac,vocos}, with all temporal receptive fields modified to depend only on current and past inputs.
The encoder maps 48~kHz mono audio to 128-dimensional continuous Gaussian latents at 100~Hz using a hop size of 480 samples, corresponding to 10~ms per frame.
The decoder contains 7.96 million parameters and consists of a causal Vocos ConvNeXt stack followed by a single inverse STFT that produces 480 waveform samples per latent frame.
Before training the LiveSynth generator, we standardize each latent dimension independently and reverse this normalization before decoding.

The VAE is trained using a weighted combination of reconstruction, adversarial, feature-matching, and KL regularization objectives:
\begin{equation}
\begin{aligned}
\mathcal{L}_{\mathrm{VAE}}
={}&
\lambda_{\mathrm{mel}}\mathcal{L}_{\mathrm{mel}}
+\lambda_{\mathrm{MSS}}\mathcal{L}_{\mathrm{MSS}}\\
&+\lambda_{\mathrm{adv}}\mathcal{L}_{\mathrm{adv}}
+\lambda_{\mathrm{FM}}\mathcal{L}_{\mathrm{FM}}
+\beta\mathcal{L}_{\mathrm{KL}}.
\end{aligned}
\end{equation}
Here, $\mathcal{L}_{\mathrm{mel}}$ sums log-mel $\ell_1$ losses across multiple scales, and $\mathcal{L}_{\mathrm{MSS}}$ combines linear- and log-magnitude $\ell_1$ losses across multiple STFT scales; $\mathcal{L}_{\mathrm{adv}}$, $\mathcal{L}_{\mathrm{FM}}$, and $\mathcal{L}_{\mathrm{KL}}$ denote the least-squares adversarial, feature-matching, and KL regularization losses, respectively.

\subsection{MIDI and Timbre Conditioning}
Streaming MIDI events are aligned to the 100~Hz latent frame rate and represented as frame-wise per-pitch states.
At each frame, each pitch is represented as inactive, offset, or onset or sustain at one of five quantized velocity levels, yielding 12 states in total.
Each active pitch--state pair is mapped to a learned embedding, and the resulting embeddings are summed across pitches.
We then add a learned baseline embedding and apply RMS normalization to reduce scale variation with the number of active notes.
For sustain states, we apply a RoPE-style rotation~\cite{rope} to the first 64 embedding dimensions according to the elapsed time since onset, leaving the remaining dimensions unchanged.
This preserves note-age information after the onset leaves the causal attention window.

We apply structured MIDI-conditioning dropout during training.
For each training clip, the entire MIDI condition is dropped with probability 0.15, a single contiguous span is dropped with probability 0.35, and the condition is retained otherwise.
The span length is sampled uniformly between one quarter and three quarters of the clip, with its starting position sampled uniformly from the valid range.
Every dropped frame is replaced with the same learned absent embedding, distinct from the baseline embedding that represents an explicitly all-inactive MIDI frame.
Full-clip dropout enables generation without MIDI conditioning, while span dropout allows the model to continue from the preceding MIDI-conditioned context.
During inference, this mechanism supports autonomous performance continuation by supplying the absent condition after a performed prefix.
The resulting frame-wise MIDI embedding is added to the projected generator input before the first Transformer block.

For timbre conditioning, we extract a 512-dimensional, L2-normalized embedding from reference audio or a text prompt using a frozen LAION-CLAP encoder~\cite{clap}.
A shared linear projection first maps the embedding to the Transformer dimension.
Each Transformer block then applies adaLN-Zero conditioning~\cite{dit}.
During training, a single audio-derived timbre embedding is shared across all frames of a clip.
At inference time, however, the conditioning interface accepts a new embedding at every frame.
LiveSynth exploits this capability for continuous timbre morphing by applying spherical linear interpolation between two CLAP embeddings over time.

Although CLAP maps audio and text into a shared space, contrastively trained multimodal encoders are known to retain a gap between modality-specific embedding distributions~\cite{mind_the_gap}.
This gap is consequential because LiveSynth is trained with audio-derived timbre conditions.
We therefore follow prior work on lightweight cross-modal embedding alignment~\cite{choi2023crossmodal} and fit an uncentered orthogonal Procrustes mapping from text to audio embeddings:\begin{equation}
\mathbf{R}^{\star}
=
\underset{\mathbf{R}^{\top}\mathbf{R}=\mathbf{I}}{\arg\min}
\left\|
\mathbf{X}\mathbf{R}-\mathbf{Y}
\right\|_{\mathrm{F}}^{2},
\qquad
\hat{\mathbf{e}}_{\mathrm{proc}}
=
\operatorname{normalize}
\left(
\mathbf{e}_{\mathrm{text}}\mathbf{R}^{\star}
\right).
\end{equation}
The rows of $\mathbf{X}$ are embeddings of template-based prompts constructed from the source and family labels of NSynth training instruments, while the corresponding rows of $\mathbf{Y}$ are their instrument-level audio embeddings.
The mapping is fitted on the NSynth training split and applied only to text conditions at inference time.

\subsection{Feedback-Free Sequence Generation}

Following LiveBand~\cite{liveband}, LiveSynth employs a feedback-free generator that produces continuous latent sequences without conditioning on its previous predictions.
At frame $t$, the generator is formulated as
\begin{equation}
\hat{\mathbf{z}}_t
=
G_{\theta}
\left(
\boldsymbol{\epsilon}_{\leq t},
\mathbf{m}_{\leq t},
\mathbf{e}_{\leq t}
\right),
\qquad
\boldsymbol{\epsilon}_t
\sim
\mathcal{N}(\mathbf{0},\mathbf{I}),
\end{equation}
where $\boldsymbol{\epsilon}_t$, $\mathbf{m}_t$, and $\mathbf{e}_t$ denote the frame-wise noise, MIDI, and timbre conditions, respectively.
Unlike autoregressive latent models, $G_{\theta}$ does not receive previously generated latents $\hat{\mathbf{z}}_{<t}$ as input.
Temporal dependencies are instead represented through the causal hidden states of the Transformer.

We use sliding-window causal attention over the most recent 500 frames, corresponding to 5~s of context.
During training, all frames are processed in parallel under the resulting band-causal attention mask.
At inference, a bounded 500-frame key--value cache provides the same causal context as the windowed attention used during parallel training, enabling one-frame-at-a-time generation.
Because generated latents are never fed back, their errors does not accumulate through subsequent inputs.

The generator is trained in the standardized VAE latent space using a conditional discriminator based on LiveBand and R3GAN~\cite{liveband,r3gan}.
The discriminator operates on standardized real or generated latent sequences.
It encodes MIDI using a separate MIDI encoder and concatenates the resulting representation with the latent input along the channel dimension.
A non-causal ConvNeXt stack then progressively downsamples the sequence while applying timbre-dependent FiLM at each block~\cite{convnext,film}.
Temporal average pooling followed by a linear head produces a single sequence-level real--fake logit.

We optimize the generator and discriminator using the paired relativistic adversarial objective of R3GAN~\cite{r3gan}.
Following LiveBand~\cite{liveband}, we also apply adaptive gradient penalties.
To reinforce adherence to the conditioning signals, we apply an $\ell_1$ latent reconstruction loss only to the fully conditioned subset $\mathcal{S}$:
\begin{equation}
\mathcal{L}_{\mathrm{rec}}
=
\frac{1}{|\mathcal{S}|dN}
\sum_{i\in\mathcal{S}}
\left\|
\hat{\mathbf{z}}^{(i)}
-
\mathbf{z}^{(i)}
\right\|_{1},
\end{equation}
where $d$ and $N$ denote the latent dimension and sequence length, respectively.
Since incomplete conditioning admits multiple valid outputs, restricting pointwise reconstruction to $\mathcal{S}$ avoids penalizing valid alternatives and encouraging averaged predictions.
The final generator objective is
\begin{equation}
\mathcal{L}_{G}
=
\mathcal{L}_{G}^{\mathrm{adv}}
+
\lambda_{\mathrm{rec}}
\mathcal{L}_{\mathrm{rec}},
\end{equation}
where $\mathcal{L}_{G}^{\mathrm{adv}}$ denotes the R3GAN generator objective.

\begin{table*}[t]
\centering
\footnotesize
\setlength{\tabcolsep}{3pt}
\begin{tabular}{l rrrrrr rrrrrr}
\toprule
& \multicolumn{6}{c}{NSynth}
& \multicolumn{6}{c}{Slakh} \\
\cmidrule(lr){2-7}
\cmidrule(lr){8-13}
Method
& MERT$\uparrow$
& MSS$\downarrow$
& PANNs$\uparrow$
& F1$\uparrow$
& CE$\uparrow$
& PQ$\uparrow$
& MERT$\uparrow$
& MSS$\downarrow$
& PANNs$\uparrow$
& F1$\uparrow$
& CE$\uparrow$
& PQ$\uparrow$ \\
\midrule
Ground truth
& 1.000 & 0.0 & 0.912 & 0.476 & 4.52 & 6.77
& 1.000 & 0.0 & 0.800 & 0.785 & 5.22 & 7.45 \\
VAE reconstruction
& 0.773 & 9.5 & 0.871 & 0.288 & 4.53 & 6.43
& 0.797 & 7.7 & 0.783 & 0.564 & 5.03 & 7.24 \\
\midrule
TokenSynth~\cite{tokensynth}
& 0.465 & 26.0 & 0.794 & 0.317 & 4.00 & \textbf{6.88}
& 0.498 & 28.7 & 0.713 & \textbf{0.391} & 3.97 & \textbf{7.10} \\
CTD~\cite{ctd}
& 0.438 & 19.5 & 0.755 & \textbf{0.361} & \textbf{4.98} & 6.78
& 0.495 & 16.1 & 0.735 & 0.339 & \textbf{5.03} & 7.08 \\
LiveSynth (ours)
& \textbf{0.540} & \textbf{13.1} & \textbf{0.819} & 0.210 & 4.16 & 5.95
& \textbf{0.506} & \textbf{12.4} & \textbf{0.743} & 0.305 & 4.21 & 6.61 \\
\midrule
AnySynth$^\dagger$~\cite{anysynth}
& 0.513 & -- & 0.820 & -- & 5.73 & 7.18
& 0.519 & -- & 0.882 & -- & 5.32 & 7.15 \\
\bottomrule
\end{tabular}
\caption{Objective evaluation of zero-shot instrument cloning on 530 NSynth and 480 Slakh clips. Ground truth and VAE reconstruction are shown as references. $^\dagger$AnySynth results are self-reported on different instrument and sample sets and averaged across multiple reference durations; they are included for context and excluded from ranking.}
\label{tab:cloning}
\end{table*}

\section{Experiments}
\label{sec:experiments}

\subsection{Experimental Setup}

\noindent\textbf{Implementation details.}
We train the causal VAE for 400k steps on 1~s, 48~kHz clips from a weighted mixture of 11 music, speech, and general-audio datasets, using a batch size of 80 and a learning rate of $10^{-4}$.
Its mel-spectrogram, multi-scale STFT, adversarial, and feature-matching losses are weighted by 15, 1, 1, and 2, respectively, with a KL weight of $10^{-3}$.

The LiveSynth generator and discriminator contain 190M and 154M parameters, respectively.
LiveSynth is trained for 200k updates on 10~s clips, comprising NSynth instruments~\cite{nsynth} rendered with Lakh MIDI~\cite{lakh} (85\%) and Slakh2100 stems~\cite{slakh} (15\%).
We upsample the original 16~kHz NSynth recordings to 48~kHz using LavaSR~\cite{lavasr} before rendering.
Each target is paired with an independently sampled 10-s reference of the same instrument or stem.
We use AdamW with a batch size of 12, a learning rate of $10^{-4}$, and $\lambda_{\mathrm{rec}}=3$.

\noindent\textbf{Evaluation setup.}
For instrument cloning, we evaluate 530 clips of 5~s each from 53 held-out NSynth instruments and 480 such clips from 12 Slakh test classes.
LiveSynth uses a 10~s reference excerpt, while TokenSynth~\cite{tokensynth} and CTD~\cite{ctd} follow their original reference-conditioning pipelines.
All systems receive the same note pitches and timings, and their outputs are resampled to 48~kHz.
Text-to-instrument synthesis is evaluated on 16 NSynth instrument classes with 32 MIDI excerpts per class and metadata-derived text prompts, using raw and Procrustes-aligned CLAP embeddings; the latter is fitted only on the training split.
We evaluate target similarity with MERT~\cite{mert}, reference-timbre similarity with PANNs~\cite{panns}, spectral reconstruction with multi-scale spectral (MSS) loss, and MIDI adherence with note F1 computed from YourMT3+ transcriptions~\cite{yourmt3}.
We assess perceptual quality using the content enjoyment (CE) and production quality (PQ) scores from Audiobox Aesthetics~\cite{audiobox}, and text--audio correspondence using MusicCoCa~\cite{live_music_models}.

\subsection{Results}

\noindent \textbf{Zero-Shot Instrument Cloning.} Table~\ref{tab:cloning} reports zero-shot instrument cloning results using reference recordings of unseen timbres.
Despite the constraints of frame-wise, strictly causal generation, LiveSynth achieves the highest MERT and PANNs similarities and the lowest MSS among the systems evaluated under the same protocol on both NSynth and Slakh.
However, LiveSynth yields lower note F1 and PQ, while its CE falls between the two offline baselines.
The VAE reconstruction results suggest that the causal audio representation contributes to this gap, particularly in MIDI adherence.

\begin{table}[t]
\centering
\footnotesize
\setlength{\tabcolsep}{4pt}
\begin{tabular}{lrrrr}
\toprule
Method & MusicCoCa$\uparrow$ & F1$\uparrow$ & CE$\uparrow$ & PQ$\uparrow$ \\
\midrule
Real instrument
& 0.152 & 0.238 & 3.78 & 6.99 \\
\midrule
TokenSynth~\cite{tokensynth}
& 0.165 & 0.202 & 3.74 & 6.97 \\
\quad w/ Procrustes
& \textbf{0.177} & 0.219 & 3.76 & \textbf{7.00} \\
\midrule
LiveSynth (ours)
& 0.149 & 0.181 & 4.27 & 6.61 \\
\quad w/ Procrustes
& 0.165 & \textbf{0.279} & \textbf{4.64} & 6.60 \\
\bottomrule
\end{tabular}
\caption{Objective evaluation of text-to-instrument.}
\label{tab:text}
\end{table}

\noindent \textbf{Text-to-Instrument.}
 Table~\ref{tab:text} evaluates text-based timbre control using raw and Procrustes-aligned CLAP embeddings.
Procrustes alignment improves MusicCoCa similarity and note F1 for both TokenSynth and LiveSynth, demonstrating the benefit of correcting the audio--text modality gap.
With alignment, LiveSynth matches the MusicCoCa similarity of unaligned TokenSynth while achieving the highest note F1 and CE among the generated outputs.
Aligned TokenSynth remains stronger than LiveSynth in text--audio correspondence and PQ, showing that prompt fidelity and production quality remain limitations of LiveSynth.

\label{sec:live}

\begin{figure}[t]
  \centering
  \includegraphics[width=0.9\columnwidth]{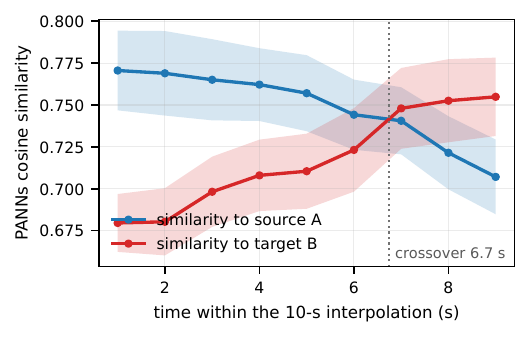}
 \caption{PANNs similarity to endpoint timbres $A$ and $B$ during a streamed 10~s timbre morph. Similarities are computed over 2~s sliding windows with a 1~s hop; curves show the mean and 95\% CI across 64 instrument pairs.}
  \label{fig:morph}
\end{figure}

\noindent \textbf{Timbre Morphing.}
We evaluate frame-wise timbre control by conditioning the generated audio on a predefined MIDI performance while spherically interpolating the timbre embedding from instrument $A$ to $B$ over a 10~s transition.
Fig.~\ref{fig:morph} shows PANNs similarity to both endpoints across 64 instrument pairs.
Similarity to $A$ decreases while similarity to $B$ increases monotonically, indicating a consistent transition between the endpoint timbres.
The curves cross after the temporal midpoint, showing that the similarity crossover does not coincide with the midpoint of the embedding interpolation.
We hypothesize that this delay arises from earlier timbre conditions retained in the Transformer's KV cache.

\label{sec:cont}

\begin{table}[t]
\centering
\footnotesize
\setlength{\tabcolsep}{5pt}
\begin{tabular}{lrrrr}
\toprule
Condition
& PCH-JSD
& IOI-JSD
& CE$\uparrow$
& PQ$\uparrow$ \\
\midrule
True continuation
& 0.251 & 0.204 & 4.73 & 6.60 \\
LiveSynth (matched)
& 0.229 & 0.529 & 4.42 & 5.68 \\
LiveSynth (shuffled pairing)
& 0.655 & 0.561 & -- & -- \\
\bottomrule
\end{tabular}
\caption{Objective evaluation of autonomous 5~s performance continuation.}
\label{tab:cont}
\end{table}

\noindent \textbf{Autonomous Performance Continuation.}
For 265 excerpts spanning 53 held-out instruments, we provide ground-truth MIDI events for a 5~s prefix and switch to the absent MIDI condition for the following 5~s while retaining the timbre condition.
From YourMT3+ transcriptions~\cite{yourmt3}, we compute the Jensen--Shannon distance between each prefix and continuation using note-count pitch-class histograms (PCH-JSD) and inter-onset-interval histograms (IOI-JSD).
The matched condition compares each prefix with its corresponding generated continuation, whereas the shuffled control compares the same prefix with a continuation generated from a different excerpt of the same instrument.
As shown in Table~\ref{tab:cont}, the matched continuations achieve PCH-JSD comparable to the true continuations and substantially lower than the shuffled control, indicating strong dependence on the preceding pitch-class distribution. 
In contrast, their IOI-JSD remains close to the shuffled control and substantially above the true continuations, indicating weak dependence on the preceding rhythmic context.
CE and PQ also remain below those of the true continuations, indicating lower perceptual quality in the generated continuations.

\noindent \textbf{Real-Time Inference.}
\label{sec:realtime}
 We evaluate real-time performance on a fanless M3 MacBook Air using the MLX int8 inference path.
Each inference step produces 480 samples (10~ms) of audio, and the strictly causal, zero-lookahead design yields an algorithmic latency of 10~ms.
Over 6{,}000 consecutive frames, LiveSynth achieves an average processing time of 8.78~ms per frame, corresponding to an RTF of 0.88, defined as processing time divided by generated audio duration.
The 99th-percentile processing time is 9.77~ms, remaining within the per-frame real-time budget.

\section{Conclusion and Limitations}

We presented LiveSynth, a streaming generative neural synthesizer for flexible timbre control and interactive generation, including zero-shot instrument cloning and text-to-instrument.
LiveSynth achieves competitive timbre similarity and lower spectral distance than offline baselines while operating at an RTF of 0.88 with 10~ms algorithmic latency, and additionally supports frame-wise timbre morphing and autonomous performance continuation.
Its limitations include lower note accuracy and audio quality, weak rhythmic dependence in continuation, and evaluation without listening tests or performer studies.
Improved causal audio representations, rhythmic modeling, and human evaluation remain future directions.

\bibliographystyle{IEEEbib}
\bibliography{refs}

\end{document}